\documentclass[12pt]{article}
\usepackage{graphicx,verbatim,array,multicol,courier,amssymb,amsmath}
\usepackage{fullpage}
\usepackage{topcapt}
\usepackage{color}
\usepackage{natbib}

\def\thick#1{\hbox{\rlap{$#1$}\kern0.25pt\rlap{$#1$}\kern0.25pt$#1$}}

\def\b0{{\boldsymbol 0}}
\def\ba{{\bf a}}

\def\bb{{\bf b}}

\def\bh{{\bf h}}
\def\bs{{\bf s}}

\def\bt{{\bf t}}
\def\bx{{\bf x}}
\def\bv{{\bf v}}
\def\bw{{\bf w}}
\def\bX{{\bf X}}
\def\by{{\bf y}}
\def\bY{{\bf Y}}
\def\bz{{\bf z}}

\def\bone{{\bf 1}}

\def\bbeta{{\thick\beta}}

\def\bpsi{{\thick\psi}}
\def\sbpsi{{\text{\footnotesize\bpsi}}}

\def\bsigma{{\thick\sigma}}
\def\bSigma{{\thick\Sigma}}

\def\sbbeta{{\text{\footnotesize\bbeta}}}

\def\bpsihat{{\widehat\bpsi}}

\def\sigmahat{{\widehat\sigma}}

\def\Dif{{\sf D}}
\def\Diff{{\sf H}}
\def\CH{\text{H}}
\def\CJ{\text{J}}
\def\CHhat{\widehat{\CH}}
\def\CJhat{\widehat{\CJ}}

\def\bpsihatmcle{{\bpsihat_{\text{\footnotesize{MCLE}}}}}
\def\Dlpsi{{\Dif_{\sbpsi}}}
\def\Hlpsi{{\Diff_{\sbpsi}}}

\def\cell{{\ell_\mathcal{C}}}
\def\pell{{\ell_\mathcal{P}}}

\begin{document}

\title{Likelihood-based inference for max-stable processes}
\author{S. A. Padoan\footnote{Laboratory of Environmental 
Fluid Mechanics and Hydrology, Ecole Polytechnique 
F\'{e}d\'{e}rale de Lausanne, Switzerland. Email: Simone.Padoan@epfl.ch},\,
M. Ribatet\footnote{Institute of Mathematics, Ecole Polytechnique 
F\'{e}d\'{e}rale de Lausanne, Switzerland. Email: Mathieu.Ribatet@epfl.ch}\:
and S. A. Sisson\footnote{School of Mathematics and Statistics, University 
of New South Wales, Australia. Email: Scott.Sisson@unsw.edu.au}
}

\maketitle

\begin{abstract}
The last decade has seen max-stable processes emerge as a common tool 
for the statistical modelling of spatial extremes.
However, their application is complicated due to the unavailability of 
the multivariate density function, and so likelihood-based methods 
remain far from providing a complete and flexible framework for inference.
In this article we develop inferentially practical, likelihood-based 
methods for fitting max-stable processes derived from a composite-likelihood approach.
The procedure is sufficiently reliable and versatile to permit the 
simultaneous  modelling of joint and marginal parameters  in the spatial 
context at a moderate computational cost. 
The utility of this methodology is examined via simulation, and illustrated by 
the analysis of U.S. precipitation extremes.
\\

\end{abstract}


\section{Introduction}
\label{sec:intro}



A common objective of spatial analysis is to quantify and 
characterise the behavior of environmental phenomena such
as precipitation levels, windspeed or daily temperatures. A number of
generic approaches to spatial modelling have been developed (e.g.
\citet{barndorff98}; \citet{cressie93}; \citet{ripley04}),
but these are not necessarily ideal for handling extremal
aspects given their focus on mean process levels. 
%
Analyses of spatial extremes are
useful devices for understanding and predicting extreme events such as 
hurricanes, storms and floods.
%
In light of recent concerns over climate change, the use of robust 
mathematical and statistical methods for such analyses has grown in importance

While the theory and statistical practice of univariate extremes is
well developed, there is much less guidance for the modelling of
spatial extremes. This is problematic as many environmental
processes 
have a natural spatial domain.
%
%
We consider a temporal series of componentwise
maxima of process measurements recorded at $k=1,\ldots,K$ 
locations, within a contiguous region. 
Observations $\{y_{n,k}\}$ each denote the maximum of $m$ samples 
over $n=1,\ldots,N$ 
temporal blocks.
For example, for daily observations, $m=366$ implies the $\{y_{n,k}\}$ describe 
process annual maxima. 


The spatial analogue of multivariate extreme value models is the class
of max-stable processes
\citep{deHaan84,deHaan+p86,resnick87}. Max-stable processes have a
similar asymptotic motivation to the univariate Generalised Extreme
Value (GEV) \citep{vonMises54,jenkinson55}), providing a
general approach to modelling process extremes incorporating temporal
or spatial dependence.
%
%
%
%
%
%
Statistical methods 
for max-stable processes and data analysis of practical problems are
discussed  by \citet{smith90}, \citet{coles93}, \citet{coles+w94} and 
\citet{coles+t96}.
Standard likelihood methods for such models are complicated by the
intractability of the multivariate density function in all but the most 
trivial cases. This presents an obstacle in the use of max-stable processes for
spatial extremes.

 There is a lack of a proper inferential framework for the
analysis of spatial extremes (although \citet{deHaan+p06} describe 
some non-parametric estimators).
In this article we develop flexible and inferentially practical methods for the
fitting of max-stable processes to spatial data based on
non-standard, {\it composite} likelihood-based methods 
\citet{lindsay88}.
%
%
An appealing feature of this approach is that the estimation of
GEV marginal parameters can be performed jointly with the 
dependence parameters in a unified framework. Accordingly, there is no need for 
separate estimation procedures. 
With highly-structured problems such as 
max-stable processes, this approach produces flexible and reliable results with 
a moderate computational cost.

The article is organised as follows:
Section \ref{sec:MaxStable} reviews the theory of max-stable processes and 
its relationship 
to spatial extremes. Our composite likelihood approach is developed in 
Section \ref{sec:MaxStableInf} and
Section \ref{sec:MaxStableSim} evaluates the method's performance 
through a number of simulation studes. We conclude with an illustration of 
a real extremal data analysis of U.S.  
precipitation levels.


\section{Max-stable processes and spatial extremes}
\label{sec:MaxStable}

\subsection{Max-stable processes}
\label{sec:MaxStableDef}

Max-stable processes provide a natural generalisation of extremal dependence 
structures in continuous spaces.
From this, closed-form bivariate distributions can be derived.
\\

\noindent \textbf{Definition}. \quad Let $T$ be an index set and 
$\{\tilde{Y}_i(t)\}_{t\in T}$, $i=1,\ldots,n$ be $n$ independent
replications of a continuous stochastic process. Assume that there
are sequences of continuous functions $a_n(t)>0$ and 
$b_n(t) \in \mathbb{R}$ such that
\[
Y(t)=\lim_{n \rightarrow\infty}\frac{\max_{i=1}^n
\tilde{Y}_i(t)-b_n(t)}{a_n(t)}, \quad t \in T.
\]
If this limit exists, the limit process $Y(t)$ is {\it a max-stable process} \citet{deHaan84}.
\\

\noindent Two properties follow from the above definition
\citet{deHaan+r77}.
Firstly, the one-dimensional marginal distributions belong to 
  the class of 
  generalised extreme value distributions (GEV), $Y \sim \text{GEV} (\mu, \lambda, \xi)$ with distribution function 
  \[
  F(y;\mu,\lambda,\xi)=\exp \left[-\left\{1+\frac{\xi(y-\mu)}
  {\lambda} \right\}_+^{-1/\xi}\right], 
  \quad -\infty<\mu,\quad\xi<\infty, \quad \lambda>0,
  \]
  where $a_+ = \max(0; a)$ and $\mu$, $\lambda$ and $\xi$ are respectively 
  location, scale and shape parameters \citet{fisher+t28}. Secondly, for any $K=2,3,\ldots$, the $K$-dimensional marginal distribution 
belongs to the class of multivariate extreme value distributions.

W.l.o.g. if $a_n(t)=n$, $b_n(t)=0$ $\forall t$, then the corresponding
process,  $\{Z(t)\}_{t\in T}$, has unit Fr\'{e}chet margins, with distribution function
$
F(z)=\exp(-1/z),  z>0.
$
This process is obtainable as standardisation of $\{Y(t)\}_{t\in T}$ through
$$
\{Z(t)\}_{t\in T}\equiv \left[\left\{1+\frac{\xi(t)(Y(t)-\mu(t))}
{\lambda(t)}\right\}_+^{1/\xi(t)}\right]_{t\in T},
$$
where $\mu(t)$, $\xi(t)$ and $\lambda(t)>0$ are now continuous functions. 
The process $Z$ is still a max-stable process.
If $Z$ is also stationary, the process may be expressed through it's {\it spectral representation} \citet{deHaan+p86}.

In detail, let $\{X_j,U_j\}_{j \geq 1}$ be  a Poisson process,
$\Pi$, on $\mathbb{R}^n \times \mathbb{R}_+$, with counting measure 
$\Pi(\cdot):=\Sigma_j \mathbb{I}_{(X_j,U_j)}(\cdot)$ and
intensity measure $\nu(dx) \times u^{-2} du$, where 
$\mathbb{I}_{(X_j,U_j)}(A)$ is the indicator function of the 
random number of points falling in a bounded set 
$A \subset \mathbb{R}^n \times \mathbb{R}_+$ and $\nu$ is
a positive measure.
For a nonnegative measurable (for fixed $t\in T$) function $f(x-t)$ 
such that
$
	\int_{\mathbb{R}^n} f(x-t) \nu(dx) = 1, \forall t \in T
$
the stochastic process
\begin{equation}\label{MaxStable}
Z(t):=\max_{j=1,2,\dots} \{U_j f(X_j-t)\}, \quad t \in T,
\end{equation}
is a stationary max-stable process \cite{deHaan84}.
\citet{smith90} reinterprets this process
in terms of environmental episodes such as storm phenomena, 
in which $U$, $X$ and 
$f$ represent respectively storm magnitude,
the center and the shape. 
\citet{schlather+t03} term this the {\it storm profile model}.

For a finite set of indexes $t_1,\ldots,t_K \in T$ and positive 
thresholds $z_1,\ldots,z_K$ for $K \in \mathbb{N}$, 
the distribution 
of the random vector $Z(t_1),\ldots,Z(t_K)$ is \cite{deHaan84}
\begin{equation}\label{DFMax-stable}
\Pr\{Z(t_k)\leq z_k, \, k=1,\ldots,K\}=
\exp\left[-\int_{\mathbb{R}^n} \max_{1 \leq k \leq K} 
\left\{\frac{f(x-t_k)}{z_k}\right\}\nu(dx)\right].
\end{equation}
It then follows that the marginal
distributions are unit Fr\'{e}chet:
$$
\Pr\{Z(t)\leq z\}=
\exp \left(- z^{-1} \int_{\mathbb{R}^n}
f(x-t)\nu(dx) \right)=\exp(-1/z).
$$
Alternative spectral representations of max-stable processes exist \citep{schlather02}.

\subsection{Extremal coefficients}
\label{sec:ExtrCoeff}

Given $n$ independent realisations of a 
random vector $Y \in \mathbb{R}^d$,
the joint distribution of componentwise maxima satisfies \citep{deHaan+r77,resnick87}
\[
\Pr\left\{\max_k \max_{j=1,\ldots,n} Y_k^{(j)}/n \leq
z\right\}= \Pr\left\{\max_{j=1,\ldots,n} Y_1^{(j)}/n \leq
z\right\}^\theta = \exp(-\theta / z), \quad z>0,
\]
for $k=1,2,\dots, K$ and common threshold $z$, 
where the rightmost term
is a Fr\'{e}chet$(\theta)$ distribution. 
The parameter $1\leq\theta\leq K$ is
the {\it extremal coefficient} and it measures the extremal 
dependence between the  margins, an important practical quantity in applications \cite{smith90}. The information
in the extremal coefficient reflects the practical number of
independent variables. 
If $K$ is finite then $\theta=1$ indicates complete dependence, whereas $\theta=K$ demonstrates  full independence.

In the max-stable process framework, from (\ref{DFMax-stable}), for all $z>0$ we have
\begin{equation*}\label{}
\Pr\{Z(t_k)\leq z, \, k=1,\ldots,K\}=
\exp(-\theta/z),
\end{equation*}
and so
\begin{equation*}\label{}
\theta = \int_{\mathbb{R}^n} \max_{1 \leq k \leq K} 
\left\{f(x-t_k)\right\}\nu(dx).
\end{equation*}
where $\theta$ again represents the effective number
of independent variables. 
\cite{schlather+t03} discuss the extremal coefficient within a max-stable context.

\subsection{Spatial models}
\label{sec:MaxStableSM}

Suppose now that $T \subseteq \mathbb{R}^2$ and that $\{X_j\}_{j \geq 1}$ are random points in $\mathbb{R}^2$. 
While for $K>2$, the general $K$-dimensional distribution function under
the max-stable process representation (\ref{MaxStable}) permits no analytically tractable form, a class of bivariate spatial models is available when 
the storm profile model, $f$,  is a bivariate Gaussian density and $\mu$ is a Lebesgue measure \citep{smith90,deHaan+p06}.
In this case, for locations $\bt_i$ and $\bt_j$ 
the bivariate  distribution function of $\{Z(\b0),Z(\bh)\}$ is
\begin{equation}\label{Gaussdist}
\begin{split}
&\Pr\{Z(\b0)\leq z_i,Z(\bh)\leq z_j\}\\
&=\exp\bigg[-\frac{1}{z_i}\Phi\left(\frac{a(\bh)}{2}+
\frac{1}{a(\bh)}\log\frac{z_j}{z_i}\right)
-\frac{1}{z_j}\Phi\left(\frac{a(\bh)}{2}+\frac{1}{a(\bh)}
\log\frac{z_i}{z_j}\right)\bigg],
\end{split}
\end{equation}
where $\bh=(\bt_j - \bt_i)^\top$, $\b0$ is the origin, $\Phi$ is the standard Gaussian distribution function, $a(\bh)=(\bh^T\bSigma^{-1} \bh)^{1/2}$
and $\bSigma$ is the covariance matrix of $f$, with covariance $\sigma_{12}$ and standard deviations $\sigma_1, \sigma_2>0$. 
A derivation of (\ref{Gaussdist}) is in Appendix A.2.
A general max-stable process with a Gaussian storm profile model, $f$, is termed a 
{\it Gaussian extreme value process} \cite{smith90}, whereas
the specific model (\ref{Gaussdist}) is the 
{\it Gaussian extreme value model} \cite{coles93}.

Second-order partial derivatives of (\ref{Gaussdist}) yield the
2-dimensional density function
\begin{equation}\label{GaussDensity}
\begin{split}
f(z_i,z_j)&=\exp\bigg\{-\frac{\Phi(w(\bh))}{z_i}
-\frac{\Phi(v(\bh))}{z_j}\bigg\}\Bigg\{
\left(\frac{\Phi(w(\bh))}{z_i^2}+\frac{\varphi(w(\bh))}{a(\bh) z_i^2}-
\frac{\varphi(v(\bh))}{a(\bh) z_iz_j}\right)\\
&\bigg(\frac{\Phi(v(\bh))}{z_j^2}+\frac{\varphi(v(\bh))}{a(\bh)z_j^2}-
\frac{\varphi(w(\bh))}{a(\bh) z_i z_j}\bigg)+
\bigg(\frac{v(\bh) \, \varphi(w(\bh))}{a(\bh)^2 z_i^2 z_j}+\frac{w(\bh) \, 
\varphi(v(\bh))}{a(\bh)^2 z_i z_j^2}\bigg)\Bigg\},
\end{split}
\end{equation}
where $\varphi$ is the standard Gaussian
density function, $w(\bh)=a(\bh)/2+\log(z_j/z_i)/a(\bh)$ and
$v(\bh)=a(\bh)-w(\bh)$. The derivation of (\ref{GaussDensity})
is in Appendix A.3.

Observe that  $a(\bh)$ measures the strength of extremal
dependence: $a(\bh)\rightarrow0$ represents
complete dependence, and (in the limit) $a(h)\rightarrow
\infty$  indicates complete independence. 
In accordance with spatial models, 
the extreme dependence 
between $Z(\b0)$ and $Z(\bh)$ decreases 
monotonically and continuously with $\bh=\| \bt_j - \bt_i \|$ \cite{deHaan+p06},
and for  fixed $\bh$ the
dependence decreases monotonically as $a(\bh)$ increases.  
Accordingly, characterisation of extremal dependence is determined by the covariance, $\bSigma$, which is therefore of interest for inference.

Due to high-dimensional distributional complexity  the study of extremal dependence  
 is commonly limited to pairwise components through the extremal coefficients
\begin{equation*}
\theta(\bh) = \int_{\mathbb{R}^2} \max\{ 
f(\bx),f(\bx-\bh)\}d\bx,
\qquad 
1\leq\theta(\bh)\leq 2.
\end{equation*}
The dependence on  $\bh$ is explicit.
Specifically for the Gaussian extreme value model,
$
\theta(\bh)=2\Phi(a(\bh)/2),
$
following an argument along the lines of Appendix A.2.
Alternative models result by considering e.g. exponential or {\it t} storm profile models \cite{deHaan+p06},
or stationary Gaussian process profile models \cite{schlather02}.

\section{Likelihood-based inference}
\label{sec:MaxStableInf}

The analysis of spatial extremes is concerned with the joint modelling of  a spatial process at large numbers of data-recording stations in a fixed region. 
As discussed in Section \ref{sec:MaxStable}, the lack of closed-form distribution for max-stable processes in greater than $K=2$ dimensions precludes straightforward use of standard maximum likelihood methods for this class of models.
We now develop inferentially practical, likelihood-based classes of max-stable processes derived from a {\it composite-likelihood} approximation \citep{lindsay88,varin08}. The procedure is sufficiently reliable and versatile to 
permit the simultaneous and consistent modelling of joint and marginal parameters in the spatial 
context at a moderate computational cost.

\subsection{Composite likelihoods}
\label{sec:MaxStableCL}

For a parametric statistical model $\mathcal{F}$ 
with density function family $\mathcal{F}=\{f(\by;\bpsi),\by$ $\in \mathcal{Y} 
\subseteq \mathbb{R}^K, \bpsi \in \Psi \subseteq \mathbb{R}^d\}$, and 
a set of marginal or conditional events 
$\{\mathcal{I}_k:k\in\mathcal{K}\}$ (for some $\mathcal{K} \subseteq  \mathbb{N}$)
subset of some sigma algebra on $\mathcal{Y}$,
the composite log-likelihood is defined by
$$
\cell (\bpsi; \by)=\sum_{k \in \mathcal{K}} \log 
f(\by \in \mathcal{I}_k;\bpsi),
$$
where $\log f(\by \in \mathcal{I}_k;\bpsi)$ is the log-likelihood 
associated with event $\mathcal{I}_k$.
First-order partial derivatives of $\cell (\bpsi; \by)$
with respect to $\bpsi$ yield
the \textit{composite score function} $\Dlpsi\cell(\bpsi;\by)$, from which 
maximum composite likelihood estimator of $\bpsi$, if unique, is obtained 
by solving $\Dlpsi\cell(\bpsihatmcle;\by)=0$.
Similarly, second-order partial derivatives of 
$\Dlpsi\cell(\bpsi;\by)$
yield the Hessian matrix 
$\Hlpsi\cell(\bpsi;\by)$  (Appendix A.1). 

The key utility of the composite log-likelihood
is it's ability, under the usual regularity conditions, to provide 
consistent and  unbiased parameter estimates when standard likelihood 
estimators are not available. 
Under appropriate conditions 
\citep{lindsay88,cox+r04} the maximum composite 
likelihood estimator is consistent and asymptotically distributed as
\[
\bpsihatmcle \, \,
\dot{\sim} \, \, \text{N}(\bpsi,\, \tilde{\text{I}}(\bpsi)^{-1})
\quad \mbox{with} \quad
\tilde{\text{I}}(\bpsi) = \CH(\bpsi) \, \CJ(\bpsi)^{-1} \,
\CH(\bpsi),
\]
where $\CH(\bpsi)=\mathbb{E}\{- \Hlpsi\cell(\bpsi;\bY)\}$ and
$\CJ(\bpsi)=\mathbb{V}\{\Dlpsi\cell(\bpsi;\bY)\}$ are
analogues of the expected information matrix and the variance of the
score vector.
Although the maximum composite likelihood estimator can be unbiased, it
may not be asymptotically efficient in that $\tilde{\text{I}}(\bpsi)^{-1}$, 
the inverse of the Godambe information matrix \cite{godambe60},
may not attain the Cram\'{e}r--Rao bound
\cite{cox+r04}.

\subsection{The pairwise setting for spatial extremes}
\label{sec:MaxStablePL}

Recall, for the spatial setting we have observations $\{y_{n,k}\}$, each 
denoting the maximum of $m$ samples over $n=1,\ldots,N$ blocks and 
$k=1,\ldots K$ locations in a continuous region. E.g. for daily observations, 
$m=366$ implies the $y_{n,k}$ describe annual maxima. Accordingly, the $K$ 
univariate marginals are approximately GEV distributed.
Despite the intractability of the multivariate max-stable process, availability 
of the bivariate form (\ref{Gaussdist}) implies  a pairwise composite 
log-likelihood may be constructed as
\begin{equation}\label{eq:MaxStablePWLik}
\pell(\bpsi;\by)=\sum_{n=1}^N \sum_{i=1}^\mathcal{K} 
\sum_{j=i+1}^{\mathcal{K}-1} \log f(y_{n,i},y_{n,j};\bpsi),
\end{equation}
where each $f(y_{n,i},y_{n,j};\bpsi)$ is a  bivariate marginal density 
based on data at locations $i$ and $j$, taken over all distinct location pairs.

In order to characterized limiting behavior by a max-stable process 
(\ref{MaxStable}) 
we require unit Fr\'{e}chet marginal distributions.
Accordingly, we consider the bijection $(Y_i,Y_j)=g(Z_i,Z_j)$ 
with inverse function given by 
\begin{equation}
\label{eq:GEVTransf}
	Z_i =\left(1 + \frac{\xi_i(Y_i-\mu_i)}{\lambda_i}\right)_+^{1/\xi_i}
	\qquad
	Z_j =\left(1 + \frac{\xi_j(Y_j-\mu_j)}{\lambda_j}\right)_+^{1/\xi_j}
\end{equation}
where $Z_i\equiv Z(t_i)$ and $Y_j \equiv Y(t_j)$,
and for each marginal $Y$, the constants $\mu$, $\xi$ and 
$\lambda>0$ ensure that $Z$ is 
unit Fr\'{e}chet distributed.
The resulting bivariate density is
\begin{equation*}\label{eq:MaxStableTransf}
f_{Y_i,Y_j}(y_i,y_j)=f_{Z_i,Z_j}\left[g^{-1}(y_i, y_j)\right]|J(y_i,y_j)|,
\end{equation*}
where $f_{Z_i,Z_j}(z_i,z_j)$ denotes the density of the Gaussian extreme 
value model (\ref{GaussDensity}), and
\begin{equation*}
|J(y_i,y_j)|=\frac{1}{\lambda_i\lambda_j}\left(1 + 
\frac{\xi_i(y_i-\mu_i)}{\lambda_i}\right)_+^{1/\xi_i-1}
\left(1 + \frac{\xi_j(y_j-\mu_j)}{\lambda_j}\right)_+^{1/\xi_j-1}.
\end{equation*}
This change of variable permits the use of GEV marginals (over unit Fr\'{e}chet) 
without reforming the problem definition. 
Hence, 
the pairwise log-likelihood (\ref{eq:MaxStablePWLik})  
allows simultaneous assessment of 
the tail dependence parameters (\ref{Gaussdist}) between pairs of sites and also the 
location, scale and shape parameters of the marginal distribution at each location. 
The parameters can not be estimated as an analytical solution of the composite score
equation. Nonetheless quasi-Newton numerical maximization routines 
(e.g. Broyden, 1967) can be applied in order to obtain maximum likelihood estimates.

Variances of parameter estimates are provided through
inverse of the Godambe information matrix, with estimates of the matrices 
$\text{H}(\bpsi)$ and $\text{J}(\bpsi)$ given by
\[
\CHhat(\bpsihatmcle)=-\sum_{n=1}^N \sum_{i=1}^\mathcal{K} 
\sum_{j=i+1}^{\mathcal{K}-1} \Hlpsi \log f(y_{n,i},y_{n,j};\bpsihatmcle)
\]
and
\[
\CJhat(\bpsihatmcle)=\sum_{n=1}^N \sum_{i=1}^\mathcal{K} 
\sum_{j=i+1}^{\mathcal{K}-1} \Dlpsi \log f(y_{n,i},y_{n,j};\bpsihatmcle) \,
\Dlpsi \log f(y_{n,i},y_{n,j};\bpsihatmcle)^\top,
\]
each evaluated at the composite maximum likelihood value.
In practice the matrix $\CHhat$ is obtained straightforwardly with 
the numerical maximization routine employed for likelihood maximization. 
An explicit expression for  $\CJhat$ is derived in Appendix A.4.

In principle, estimating unique marginal parameters for each location 
ensures correct model application by respecting marginal constraints, 
though computational issues arise for large numbers of parameters.
Alternatively, as is common in the modelling of univariate extremes, 
we may describe the GEV parameters through parsimonious regression models, 
which may be functions of space, environmental and other covariates and random effects.
Specifically,  we may express each parameter as
\begin{equation}\label{eq:Predict}
\eta(\bx) \equiv h\{f(\bx)\} = \bX \bbeta,
\end{equation}
where $h$ is a link function, $\bx=(x_1,\ldots,x_d)$ is a vector of predictors, 
$\bX$ is a $N \times (d+1)$ 
design matrix, 
and $\bbeta$ is a $(d+1)\times 1$ vector of unknown parameters.

For further flexibility, a non-parametric approach may provide a useful alternative.
Non-parametric modelling of univariate extreme value responses has been recently 
proposed by \cite{chavez+d05}, \cite{yee+s07}
and  \cite{padoan+w08}. 
The work of \cite{kammann+w03}, who developed a non-parametric spatial regression with 
Gaussian response, may be extended to the current spatial extremes setting.

\subsection{Model selection}
\label{sec:MaxStableModSel}

There are two model selection approaches under the composite likelihood framework.
For nested models \cite{varin08} 
the $p$-dimensional parameter, 
$\bpsi$ is partitioned as $\bpsi=(\bpsi',\bpsi'')$, where $\bpsi'$ is 
$q$-dimensional,
and
testing $\bpsi'=\bpsi_0$ versus a two-sided hypothesis proceeds via the
composite likelihood ratio test 
statistic
\[
	W(\bpsi_0)=2\{\ell_{\mathcal{C}}(\bpsihat)-
	\ell_{\mathcal{C}}(\bpsi_0,\bpsihat''(\bpsi_0))\}. 
\]
Under the null \cite{kent82}
\begin{equation}\label{eq:likRatioStat}
W \, \dot{\sim} \, \sum_{j=1}^q \nu_i \chi_i^2,
\end{equation}
where $\chi_i^2$ are independent $\chi^2_1$ random variables, and $\nu_1\geq \ldots \geq \nu_q$ 
are the eigenvalues of
$(\CH^{\sbpsi'\sbpsi'})^{-1} \, \tilde{\text{I}}_{\sbpsi'\sbpsi'}$.
Here, $\CH^{\sbpsi'\sbpsi'}$ and and $\tilde{\text{I}}_{\sbpsi'\sbpsi'}$ respectively denote
the information matrix and the Godambe information matrix, each restricted to those elements
associated with parameter $\bpsi'$.
(Dependence on $\bpsi$ under the null is omitted for brevity.)
Hypothesis testing based on
(\ref{eq:likRatioStat}) either approximates null distribution using
estimates of the eigenvalues $\nu_i$ \cite{rotnitzky+j90},
or adjusts the composite likelihood  
such that the usual asymptotic $\chi^2_q$ null is preserved \cite{chandler+b07}.

The {\it composite likelihood information
criterion}  (CLIC) \cite{varin+v05}, useful in the 
case of non-nested models, performs model selection on the basis 
of expected Kullback--Leibler divergence between the true unknown 
model and the adopted model \citep[p.~123]{davison03}. 
In the composite likelihood context, this is the AIC under model mis-specification 
\citep{takeuchi76}; \citep[p.~150--152]{davison03}.
Model selection is based on the model minimising   
$$
-2 \{\ell_{\mathcal{C}}(\bpsihatmcle;\bY) -
\text{tr}[\CJhat(\bpsihatmcle) \, 
\CHhat(\bpsihatmcle)^{-1}]\},
$$
where the second term is the usual composite log-likelihood penalty term.

\section{Simulation Study}
\label{sec:MaxStableSim}

\begin{figure}[h]
  \centering
  \includegraphics[angle=-90,width=\textwidth]{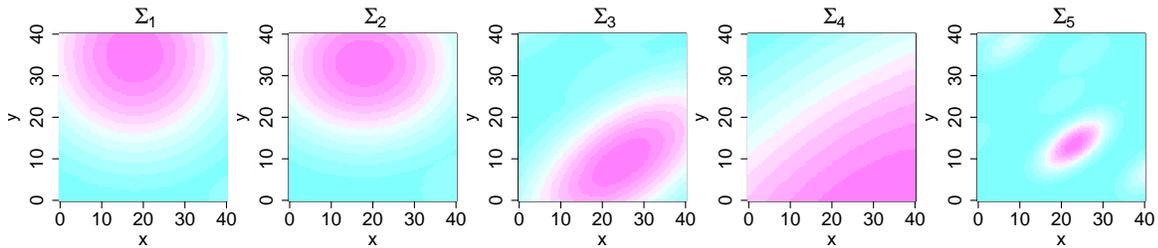}
  \caption{{\it\small A Gaussian extreme value process realisation for
    $\bSigma_1,\ldots,\bSigma_5$.
    Max-stable process simulation routines are available in the {\tt RandomFields} package in {\tt R} (Schlather, 2002). 
    }}
  \label{fig:realGaussExtValProc}
\end{figure}

We now evaluate the utility of the composite likelihood in the spatial extremes context.
We examine various forms of extremal dependence with the Gaussian storm
profile (\ref{GaussDensity}) characterised through the covariance, $\bSigma$, including 
directional and strength of dependence variations (Table \ref{SpatConf} and  Figure 
\ref{fig:realGaussExtValProc}). The covariance has direct meteorological interpretation 
and defines the extremal dependence directly.
The $K$ site locations are uniformly generated over a $40\times 40$ region. 
Given the moderate computational demand  for large site numbers, likelihood maximisation 
(and other) routines have been implemented in  {\tt C} and collected in the forthcoming {\tt R} 
package {\tt SpatialExtremes}.

\begin{table}[h]
\begin{center}
\topcaption[Extremal dependence configurations]{{\it\small Extremal dependence configurations.
}\label{SpatConf}}
\begin{tabular}{clccc}
\hline
& Spatial dependence structure & $\sigma^2_{1~}$ & $\sigma^2_{2~}$ &
$\sigma_{12}$\\
\hline
$\bSigma_1$:  &Same strength in both directions& 300 & 300 & 0\\
$\bSigma_2$:  &Different strength in both directions & 200 & 300 & 0\\
$\bSigma_3$:  &Spatial correlation & 200 & 300 & 150\\
$\bSigma_4$:  &Strong dependence & 2000 & 3000 & 1500\\
$\bSigma_5$:  &Weak dependence & 20 & 30 & 15\\
\hline
\end{tabular}
\end{center}
\end{table}
\begin{table}[htb]
  {\small
    \centering
    \topcaption
    {{\it\small Composite MLE's  based
      on 500 spatial extreme data simulations  ($K=50$ sites and $N=100$) with the Gaussian extreme value model. True values
      are in [brackets]. Standard errors obtained through Godambe
      estimates and sample standard deviation (in parantheses)}.}
    \label{compSpatial}
    \begin{tabular}{lcccccc}
      \hline
      & $\sigmahat^2_{1~}$ / s.e. & $\sigmahat^2_{2~}$ / s.e. &
      $\sigmahat^2_{12}$ / s.e.\\
      \hline
      $\bSigma_1$:  &306 [300] / 40.6 (44.7) & 306 [300] / 39.8 (41.5) &
      1 [0] / 27.9 (27.7)\\  
      $\bSigma_2$:  &204 [200] / 26.7 (28.5) & 305 [300] / 39.6 (39.7) &
      1 [0] / 21.9 (21.2)\\
      $\bSigma_3$:  &202 [200] / 25.1 (26.1) & 300 [300] / 37.3 (37.9) &
      150 [150] / 25.5 (26.1)\\
      $\bSigma_4$:  & 2053 [2000] / 495.2 (300.1) & 3065 [3000] /
      664.8 (483.1) & 1550 [1500] / 412.0 (322.4)\\
      $\bSigma_5$:  &20 [20] / 1.5 (1.6) & 30 [30] / 2.3 (2.3) & 15
      [15] / 1.6 (1.6)\\
      \hline
    \end{tabular}
  }
\end{table}

Table \ref{compSpatial} summarizes estimator performance based on moderately 
sized datasets: $K=50$ sites and $N=100$ observations.
Estimate means and standard errors over 500 data replications are reported, indicating 
good correspondence to the true values.
There is no evidence of bias, even in cases where poorer performance may be expected 
such as very strong or weak dependence.
Overall, the Godambe standard errors and sample standard deviations are consistent, 
though there is some discrepancy in the case of strong dependence.
Here, the eigenvalues of  $\bSigma_4^{-1}$ are $5 \pm \sqrt{10} / 7500$
and so we are near the parameter space boundary.
 Consequently, for some of the 500
replications, the asymptotic normality of $\bpsihatmcle$ fails, and so the Godambe 
standard errors are not relevant.
\begin{table}[h]
    \centering
  \topcaption{{\it\small 
    Normalised mean squared error of extremal coefficient estimates based on 500 data 
    simulations with $K=50$ sites and $N=100$ observations.
    Estimators are the composite MLE  and those proposed by Smith (1990) and Schlather and Tawn (2003).
    Standard deviations are reported in parantheses. The values' order of magnitude is
    $10^{-4}$}.}
  \label{tab:extCoeff}
  \begin{tabular}{lccc}
    \hline
    & Composite MLE & Smith & Schlather \& Tawn\\
    \hline
    $\bSigma_1$: & 3.1 (~5.4) & 18.4 (26.7) & 17.3
    (29.9)\\
    $\bSigma_2$: & 2.9 (~4.8) & 19.1 (30.3) & 18.9
    (28.6)\\
    $\bSigma_3$: & 2.5 (~4.7) & 21.9 (35.7) & 21.1
    (32.4)\\
    $\bSigma_4$: & 3.0 (35.6) & 10.9 (18.9) & ~6.7
    (13.3)\\
    $\bSigma_5$: & 0.3 (~0.8) & 30.3 (44.6) & 25.5
    (40.7)\\
    \hline
  \end{tabular}
\end{table}

Table~\ref{tab:extCoeff} depicts normalised mean squared errors
for three different estimators of the extremal coefficient functions:
the composite MLE derived from the Gaussian extreme value model, and those
proposed by \cite{smith90} and \cite{schlather+t03}.
Normalized mean square errors are used to prevent
the largest extremal coefficients from dominating. From Table \ref{tab:extCoeff}, in general it 
is clear that  the composite likelihood estimator is the most accurate. 
The standard deviation in the strong dependence case again suffers from the failure of the asymptotic normality of  $\bpsihatmcle$.

\begin{table}[h]
\centering
\topcaption{{\it\small
Composite MLEs for a varying number of sites ($K$) and observations ($N$), based on 500 simulations of spatial
extreme data using the Gaussian extreme value model with covariance
 $\bSigma_3$ (Table \ref{SpatConf}). 
Standard deviations are reported in parantheses}.
}\label{CMLesample}
\begin{tabular}{ccccccc}
\hline
 & $K$ & $N=10$ & $N=50$ & $N=100$ & $N=500$ & True \\
\hline
$\sigmahat^2_{1~}$ &10&245 (120.2)&207 (43.8)&205 (31.7)&199 (13.3)&200\\
 &50&244 (~90.4)&208 (37.5)&200 (28.3)&199 (11.4)&\\
 &100&239 (~94.3)&205 (37.8)&202 (30.4)&199 (11.5)&\\
\hline
$\sigmahat^2_{2~}$ &10&353 (159.1)&305 (63.9)&301 (44.8)&298 (19.5)&300\\
 &50&353 (131.6)&309 (56.6)&303 (44.3)&298 (16.9)&\\
 &100&361 (143.4)&307 (59.5)&301 (44.8)&299 (16.7)&\\
\hline
$\sigmahat_{12}$ &10&174 (108.9)&153 (41.8)&151 (31.9)&149 (13.4)&150\\
&50&179 (~91.2)&156 (38.8)&149 (28.6)&149 (11.4)&\\
 &100&181 (100.9)&154 (39.2)&150 (29.7)&150 (11.2)&\\
\hline
\end{tabular}
\end{table}

Estimator performance for a range of dataset sizes ($N=10,50,100,500$) and site numbers ($K=10,50,100$) is listed in Table \ref{CMLesample}, under
500 data replications of spatial dependence model $\bSigma_3$.
As expected, the simulations indicate that there is some bias and larger variance for small $N$, and negligible bias and small variance for large $N$.
Observe that, for fixed sample size, the number of
sites does not impact the estimation results.
\begin{figure}[h]
  \centering
  \includegraphics[width=5.4in]{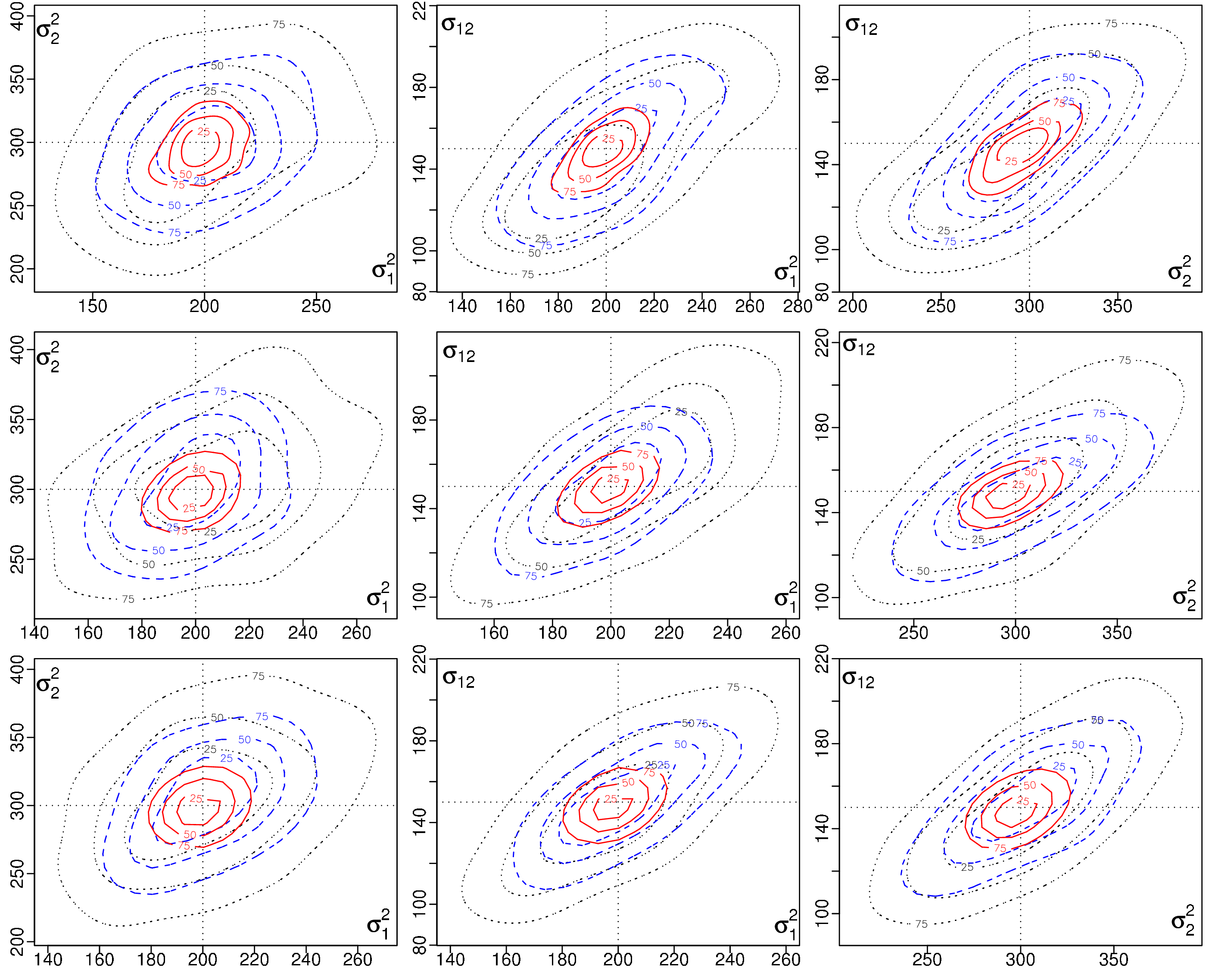}
  \caption{{\it\small Composite MLE distribution estimates for $N=50$ (dotted line), $N=100$ (broken line) and $N=1000$ (solid line) observations over $K=10$ (top row), $K=50$ (centre row) and $K=100$ (bottom row) sites. Contours correspond to 0.25, 0.5 and 0.75 percentiles.  
    }}
  \label{SimResPlots}
\end{figure}
This behavior is illustrated in Figure \ref{SimResPlots}, 
highlighting, in particular, good estimator performance with increasing $N$.

\begin{figure}[h]
  \centering
  \includegraphics[angle=-90,width=5.5in]{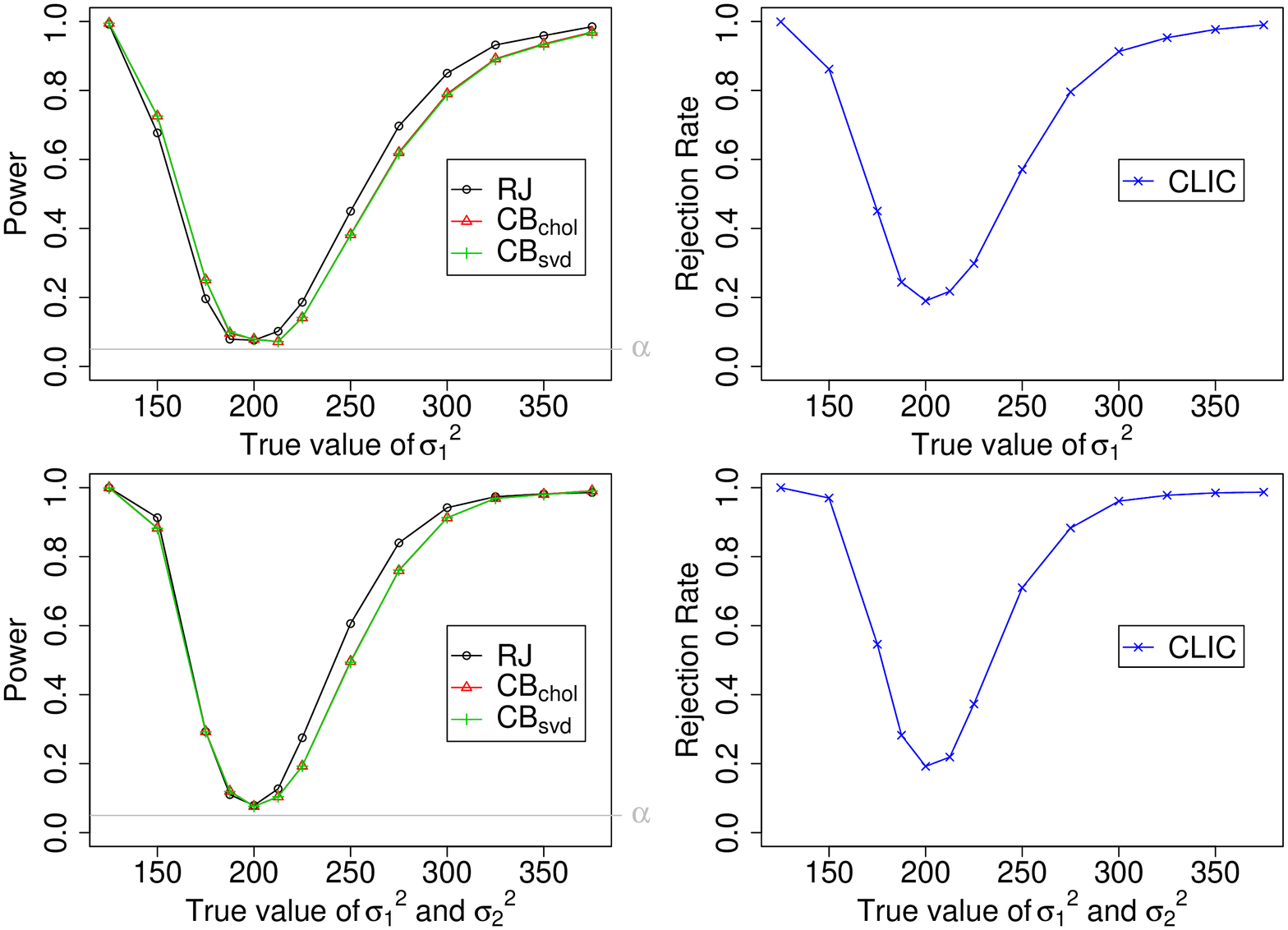}
  \caption{{\it\small Left panels: Power curves for the composite likelihood ratio
    tests; $RJ$ (Rotnitzki and Jewell, 1990), and  $CB_{chol}$ and $CB_{svd}$ (Chandler and Bate, 2007) 
    using Cholesky and singular value decompositions. Right panels: CLIC rejection rates. Test levels are $\alpha=0.05$. Point estimates are based on 1000 data replications.
    }}
  \label{fig:powerCurves}
\end{figure}

Finally, we consider model selection under misspecification. 
Figure~\ref{fig:powerCurves} illustrates power curves (likelihood ratio tests)
and rejection rates (CLIC) for two hypotheses,  each versus their complement. Namely, $H_0: \sigma_1^2 = 200$ fixing $\sigma_2^2=300$  and $\sigma_{12} = 150$ (top plots),  and $H_0: \sigma_1^2 = \sigma_2^2 = 200$ fixing $\sigma_{12} = 0$ (bottom plots).
All resulting curves are near-quadratic and, for the pairwise likelihood ratio based tests, rejection rates are close to the confidence level
$\alpha=0.05$ when the null hypothesis is true.
In this study, 
contrary to the results derived by 
\cite{chandler+b07}, the adjustment of the 
$W$ statistic (\ref{eq:likRatioStat})  \cite{rotnitzky+j90} appears to have slightly more
power -- even when testing mutiple parameters. 
In contrast the
CLIC statistic demonstrates poor performance, with the rejection rate under the
true model reaching only $20\%$.

\section{Application to U.S. precipitation data}
\label{sec:MaxStableRDA}

\begin{figure}
  \centering
  \includegraphics[angle=-90,width=.75\textwidth]{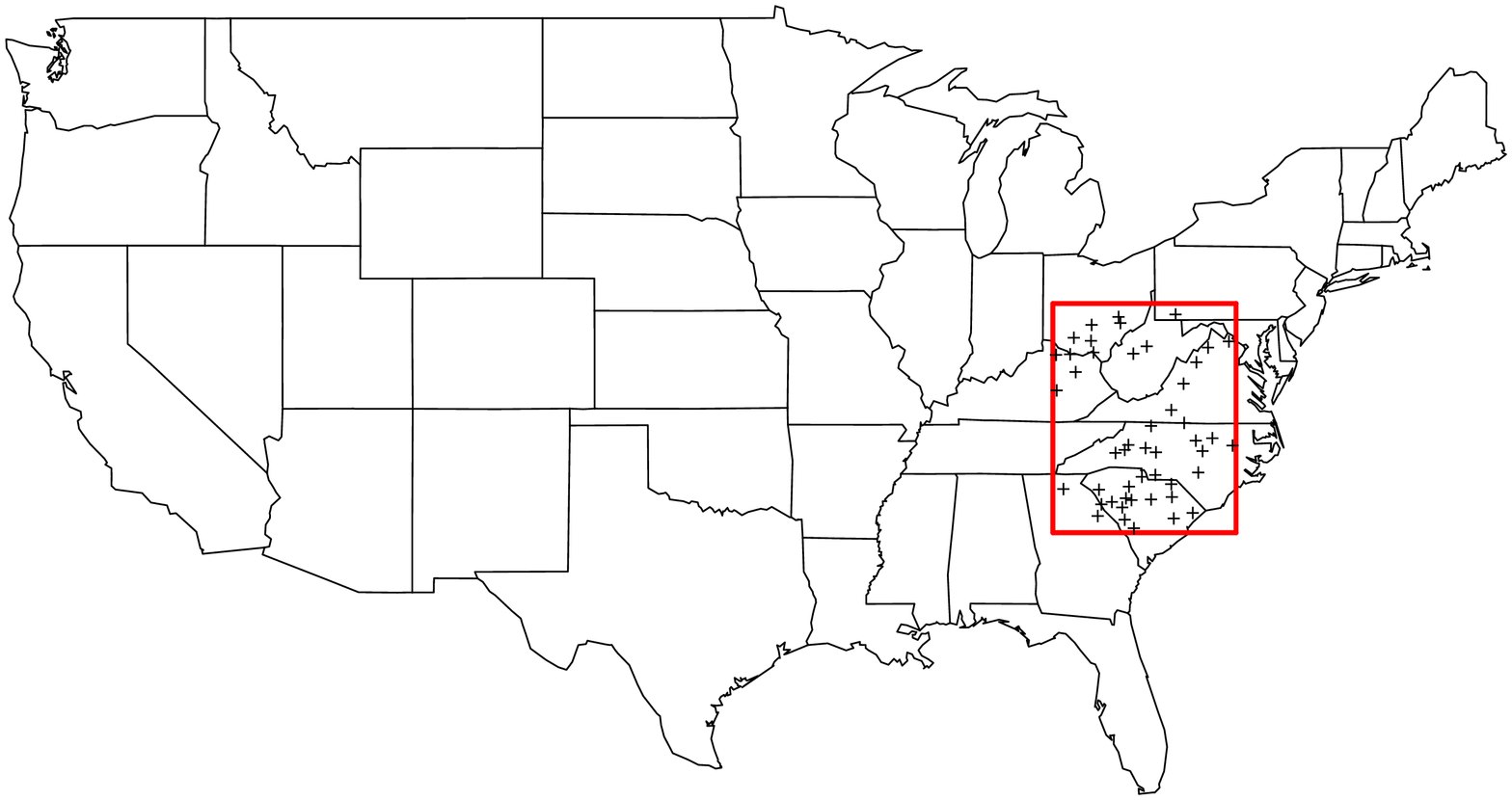}
  \caption{{\it\small Locations of the $46$ gauging stations}.}
  \label{fig:regionStudy}
\end{figure}
We illustrate the developed methodology in an analysis of U.S. precipitation data. These
data consist of $46$ gauging stations  with daily rainfall records over a period
of $91$ years (Figure~\ref{fig:regionStudy}). 
We express the model GEV parameters as simple linear functions of
longitude, lattitude and altitude. Further model variations, including
additional environmental covariates, and regressions of the covariance
$\bSigma$ on these covariates to examine spatial variations in
extremal dependence were not considered.

Exploratory analyses of the functional complexity of the GEV
parameters was performed by fitting independent GEV models to the data
from each station and evaluating appropriate surface responses using
standard techniques (e.g. ANOVA, Fisher tests).
After identifying an upper limit to model complexity, we perform model
selection on the pairwise composite likelihood of the Gaussian extreme
value process using the methodologies outlined in
Section~\ref{sec:MaxStableModSel}.

\begin{table}
  \centering
  \caption{{\it\small Some Gaussian extreme value processes
    and their corresponding maximised negative composite log-likelihood,
    degrees of freedom and the CLIC score.}}
  \label{tab:modelUS}
  \begin{tabular}{llccc}
    \hline 
    Model & & $-\pell(\bpsihatmcle;\by)$ & d.f. & CLIC\\
    \hline
    $M_0:$ & $\mu(x) = \alpha_0 + \alpha_1(\mbox{lat}) +
    \alpha_2(\mbox{alt}) + \alpha_3(\mbox{lon})$ & 412,110.2 & 12 &
    12,848,229\\ 
    & $\sigma(x) = \beta_0 + \beta_1(\mbox{lat}) +
    \beta_2(\mbox{alt}) + \beta_2(\mbox{lon})$ & & &\\
    \vspace{0.2cm}
    & $\xi(x) = \gamma_0$ & & &\\
    $M_1:$ & $\mu(x) = \alpha_0 + \alpha_1(\mbox{lat}) +
    \alpha_2(\mbox{alt})$ & 412,110.9 & 11 & 16,096,068\\
    & $\sigma(x) = \beta_0 + \beta_1(\mbox{lat}) +
    \beta_2(\mbox{alt}) + \beta_3(\mbox{lon})$ & & &\\
     \vspace{0.2cm}
    & $\xi(x) = \gamma_0$ & & &\\
    $M_2:$ & $\mu(x) = \alpha_0 + \alpha_1(\mbox{lat}) +
    \alpha_2(\mbox{alt}) + \alpha_3(\mbox{lon})$ & 412,113.3 & 11 &
    1,008,997\\
    &$\sigma(x) = \beta_0 + \beta_1(\mbox{lat}) + \beta_2(\mbox{alt})$
    & & &\\
     \vspace{0.2cm}
    & $\xi(x) = \gamma_0$ & & &\\
    $M_3:$ & $\mu(x) = \alpha_0 + \alpha_1(\mbox{lat}) +
    \alpha_3(\mbox{lon})$ & 412,234.1 & 11 & 1,926,389,242\\
    &$\sigma(x) = \beta_0 + \beta_1(\mbox{lat}) + \beta_2(\mbox{alt})
    + \beta_3(\mbox{lon})$ & & & \\
     \vspace{0.2cm}
    & $\xi(x) = \gamma_0$ & & &\\
    $M_4:$ & $\mu(x) = \alpha_0 + \alpha_1(\mbox{lat}) +
    \alpha_2(\mbox{alt}) + \alpha_3(\mbox{lon})$ & 412,380.5 & 11 &
    33,209,042\\
    &$\sigma(x) = \beta_0 + \beta_1(\mbox{lat}) + \beta_3(\mbox{lon})$
    & & & \\
     \vspace{0.2cm}
    & $\xi(x) = \gamma_0$ & & &\\
    $\mathbf{M_5:}$ & $\mathbf{\mu(x) = \alpha_0 +
      \alpha_1(\mbox{lat}) + \alpha_2(\mbox{alt})}$ &
    \textbf{412,113.3} & \textbf{10} & \textbf{1,008,261}\\
    &$\mathbf{\mathbf{\sigma(x) = \beta_0 + \beta_1(\mbox{lat}) +
        \beta_2(\mbox{alt})}}$ & & & \\
         \vspace{0.2cm}
    & $\mathbf{\xi(x) = \gamma_0}$ & & &\\
    $M_6:$ & $\mu(x) = \alpha_0 + \alpha_1(\mbox{lat})$ &
    412,237.3 & 9 & 1,086,347\\
    &$\sigma(x) = \beta_0 + \beta_1(\mbox{lat}) + \beta_2(\mbox{alt})$
    & & & \\
    & $\xi(x) = \gamma_0$ & & &\\
    \hline
  \end{tabular}  
\end{table}

Table~\ref{tab:modelUS} summarizes some of the different models
investigated. 
According to the CLIC criterion, 
model $M_5$ (in bold) is the preferred choice, although models $M_2$ and $M_6$ also appear
competitive. Despite the relatively poor performance of the CLIC criterion for model selection 
(see Section~\ref{sec:MaxStableSim}), the same
conclusions are obtained from the deviance based tests, with
$p$-values around 0.9. 
For model $M_5$, the
covariance $\bSigma$ is estimated as $\sigma_1^2 = 0.06323$
$(0.06463)$, $\sigma_{12} = 0.01334$ $(0.03661)$ and $\sigma_2^2 =
0.02581$ $(0.02538)$, 
where the parantheses indicate standard errors.

%
%

\begin{figure}
  \centering
  \includegraphics[angle=-90,width=.5\textwidth]{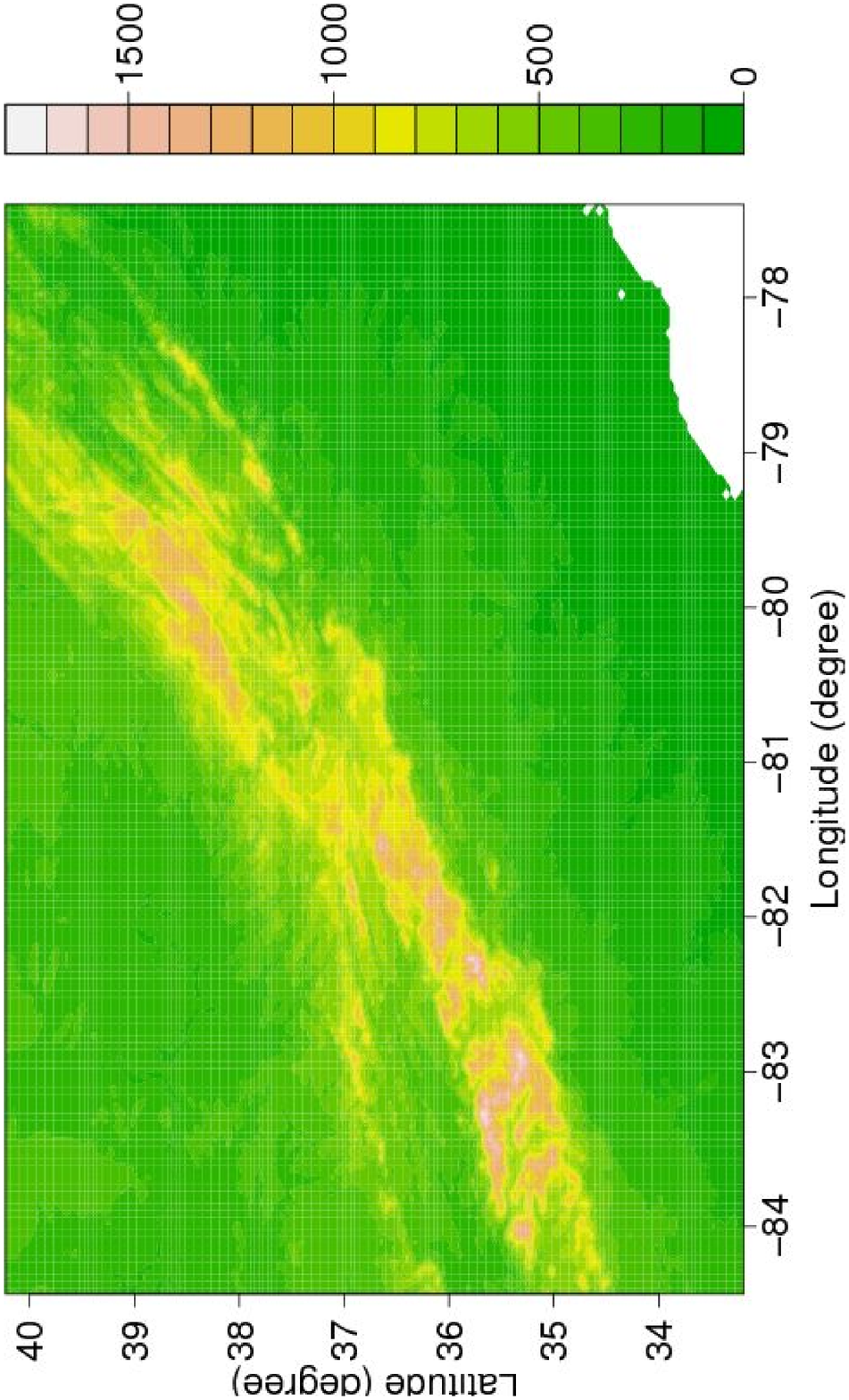}\hfill%
  \includegraphics[angle=-90,width=.5\textwidth]{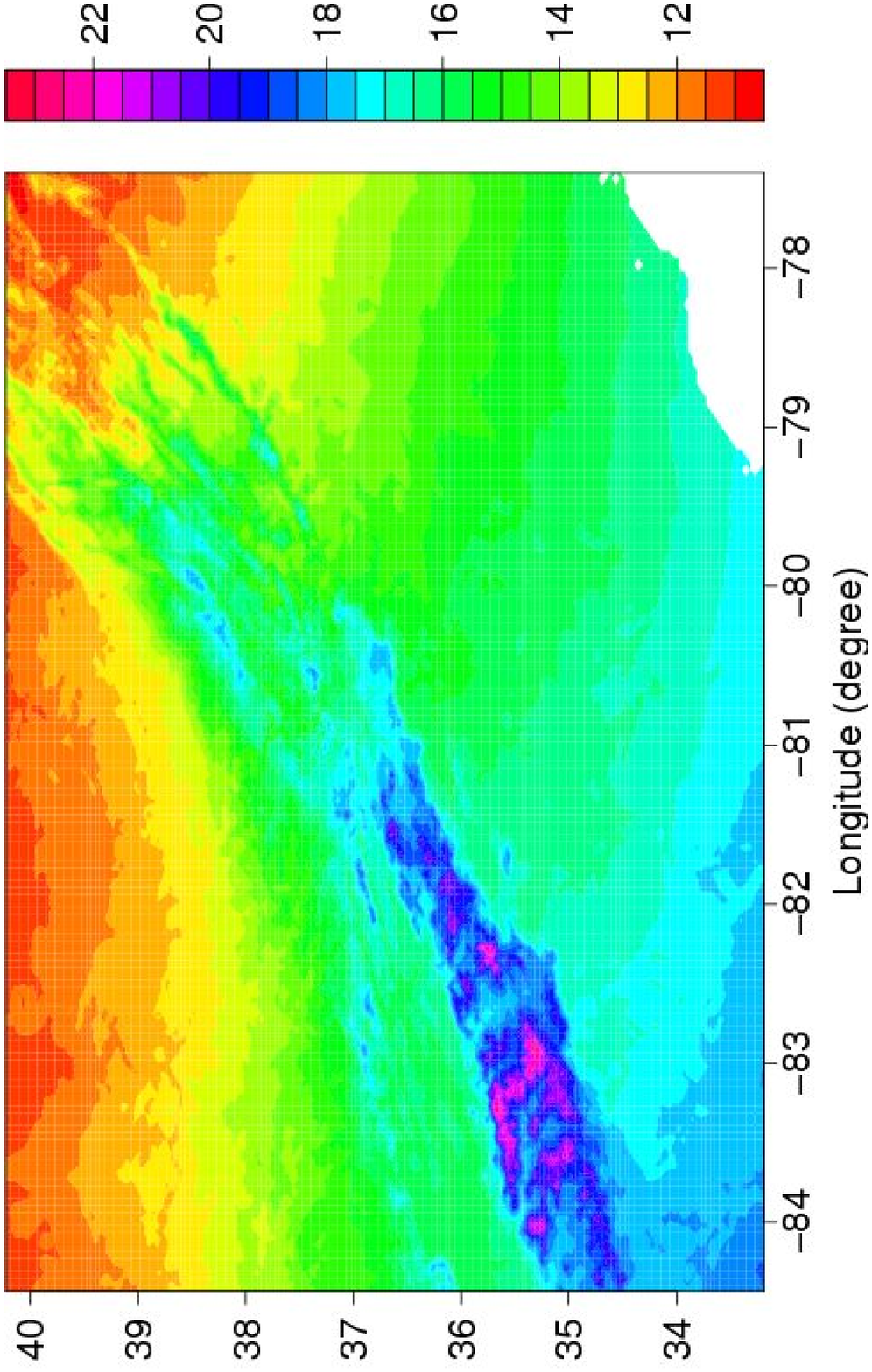}
  \caption{{\it Left: Elevation map (metres) of the region;
    Right: Pointwise 50-year return level map (cm) estimated from the
    fitted Gaussian extreme value process}.}
  \label{fig:elev+retlev}
\end{figure}

\begin{figure}
  \centering
  \includegraphics[angle=-90,width=.49\textwidth]{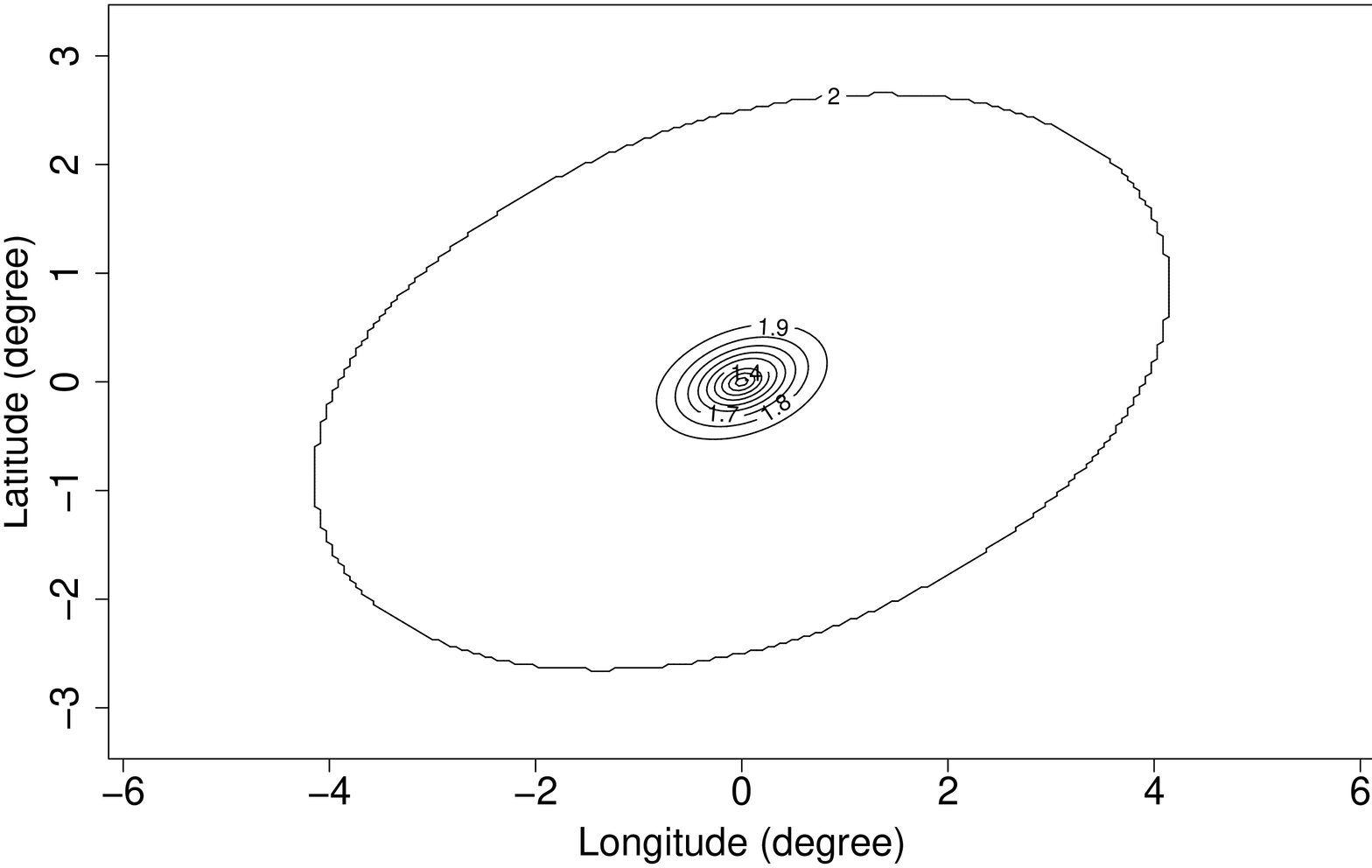}\hfill%
  \includegraphics[angle=-90,width=.5\textwidth]{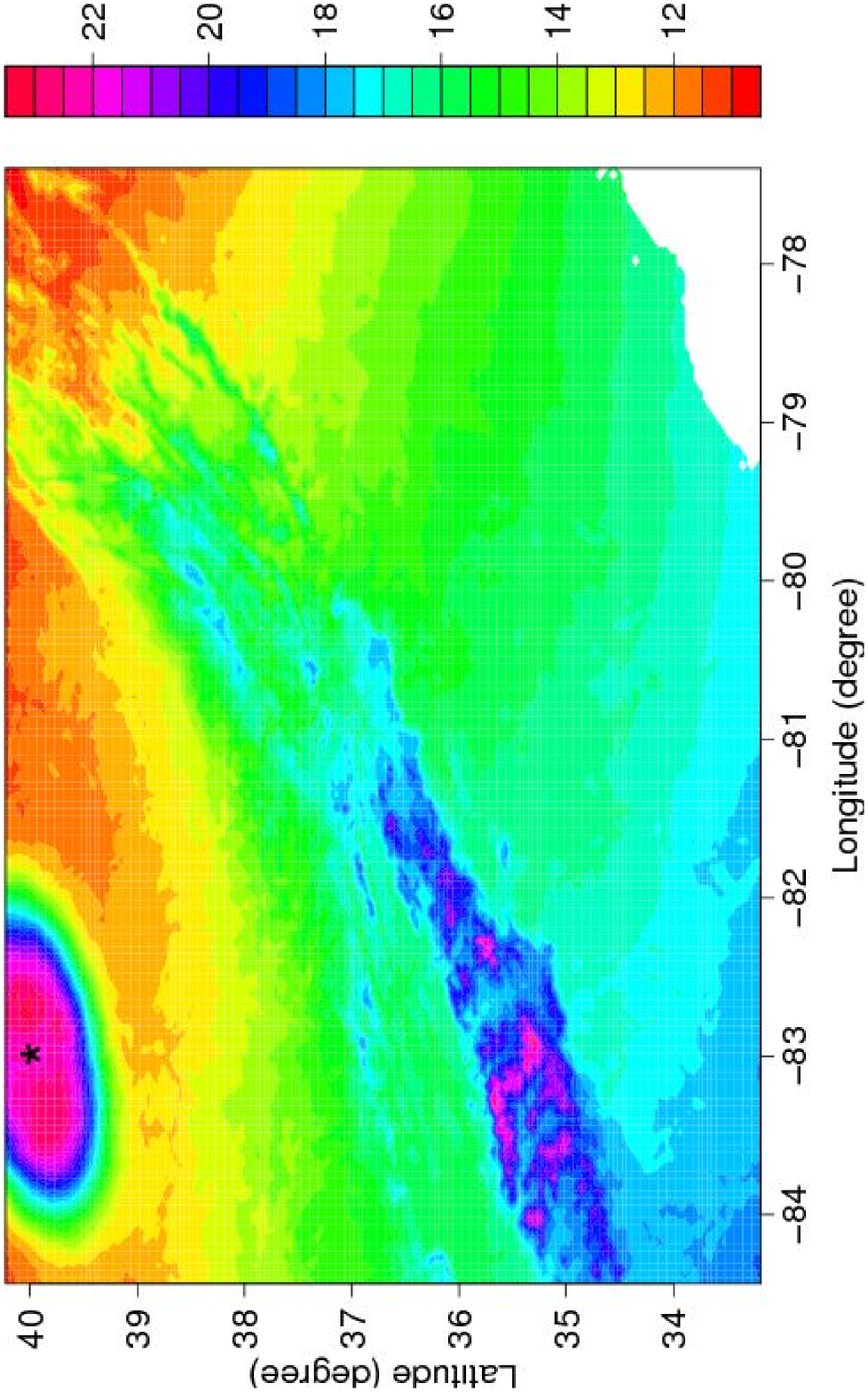}
  \caption{{\it
    Left: Contour plot of the extremal coefficient;
    Right: Pointwise 50-year, conditional return level map (cm) estimated from the
    fitted Gaussian extreme value process. The star indicates the fixed site
    used in the computation of the conditional return levels}.}
  \label{fig:extc+condQ}
\end{figure}

Figure~\ref{fig:elev+retlev} (right plot) illustrates the spatial variation of
pointwise 50-year return level estimates.
Comparison to the regional elevation map  (left plot)
indicates that the most extreme precipitation events
occur in mountainous regions.
Figure~\ref{fig:extc+condQ} (left plot) also depicts the strength of spatial dependence through
the extremal coefficient function. 
There is clear evidence of anisotropy, with 
stronger dependence in the north-east/south-west
direction. Interestingly, this axis corresponds to the
shape of the Appalachian Mountains as well as the coastline. 
Accordingly, this  directional extremal dependence may be the consequence of
storms following either the coastline or the
massif.
Finally, conditioning of a fixed site (and for a given threshold)
 by using the pairwise conditional distribution, the conditional 
$q$-year return level estimates can easily be provided. 
Figure~\ref{fig:extc+condQ} (right plot) illustrates 
the spatial variation of conditional pointwise 50-year return level estimates,
given the fixed site indicated by the star.

\section{Conclusion}
\label{sec:Conclusion}

As a natural generalisation of extremal dependence structures, max-stable processes are a powerful tool for the modelling of multivariate extremes. Unfortunately, the intractability of the multivariate density function precludes inference except in trivial cases (e.g. bivariate), or requires additional approximations and immense  computational overheads \citep{jiang+t04,sisson+ft07,peters+fs08}.

This article has developed composite likelihood-based
inferential methods for general max-stable processes. Our results demonstrate good applicability in the spatial context.
The benefits of this likelihood-based approach are the flexible joint modelling of marginal and dependence parameters,  coupled with
good estimator behaviour 
with finite samples, all
at moderate computational cost.

Modifications of the model formulation would draw alternative representations of extremal modelling
into the composite-likelihood based framework, given the known links between these and  block maxima (GEV) approaches (e.g. \cite{coles01}). 
These include threshold excess models for marginals 
(Davison and Smith, 1990), and the limiting Poisson characterisation
of extremes.  
The obvious practical benefit from these extensions would be the incorporation of more data
into the modelling process.

%

\section*{Acknowledgements}

This work was commenced while SAP was visiting the School
of Mathematics and Statistics, University of New South Wales,
Sydney, Australia. Their hospitality is gratefully acknowledged.
The authors are grateful to Anthony Davison and Stuart Coles
for their helpful suggestions, and to Richard 
Smith for providing the U.S. precipitation data.
The work was supported by the CCES Extremes project,
\\ \texttt{http://www.cces.ethz.ch/ projects/hazri/EXTREMES}.
SAS is supported by the Australian Research Council through the
Discovery Project scheme (DP0877432). 

\bibliographystyle{apalike}
\bibliography{max-stable}

\newpage

\section*{Appendix}
\label{sec:appen}

We present explicit expressions for the distribution
function (\ref{Gaussdist}), density function (\ref{GaussDensity})
and the derivatives required for the estimated covariance matrix
in Section \ref{sec:MaxStablePL}

\subsection*{A.1: Vector notation}

Let $f$ be a real-valued function in the $d\times1$ vector
$\bx=(x_1,\ldots,x_d)$. Then the $1\times d$ derivative vector, ${\sf
D}_{\bx}f(\bx)$, has $i$-th element $\partial
f(\bx)/\partial x_i$. The corresponding Hessian matrix is given by
${\sf H}_{\bx}f(\bx)={\sf D}_{\bx}\{{\sf D}_{\bx}f(\bx)\}^\top$.

If $\ba=(a_1,\ldots,a_d)$ and $\bb=(b_1,\ldots,b_d)$ are two
$d\times1$ vectors, then element-wise multiplication is denoted by
$\ba\odot\bb=(a_1b_1,\ldots,a_db_d)$. The expression $\ba/\bb$
denotes element-wise division $(a_1/b_1,\ldots,a_d/b_d)$. Scalar
functions applied to vectors are also evaluated element-wise. For
example, $\ba^{-1/\xi}=(a_1^{-1/\xi},\ldots,a_d^{-1/\xi})$.

\subsection*{A.2: Derivation of the bivariate  distribution  function}



In order to derive the cumulative distribution function (\ref{Gaussdist}), considering
formula (\ref{DFMax-stable}) and the assumptions of Section \ref{sec:MaxStableSM}, we need to solve:
\begin{equation*}\label{}
\begin{split}
F(z_i,z_j)&=\exp\left\{- 
\int_{-\infty}^\infty \int_{-\infty}^\infty
\max \left(\frac{f(x_1,x_2)}{z_i}, \frac{f(x_1-t_1,x_2-t_2)}{z_j}
\right) \, \mbox{d}  x_1 \, \mbox{d}  x_2\right\}\\
&=\exp\Bigg\{- 
\int_{-\infty}^\infty \int_{-\infty}^\infty
\frac{f(x_1,x_2)}{z_i} \mathbb{I}\left(
\frac{f(x_1,x_2)}{z_i} \geq \frac{f(x_1-t_1,x_2-t_2)}{z_j}
\right) \, \mbox{d} x_1 \, \mbox{d} x_2\\
&-\int_{-\infty}^\infty \int_{-\infty}^\infty
\frac{f(x_1-t_1,x_2-t_2)}{z_j} \mathbb{I}\left(
\frac{f(x_1-t_1,x_2-t_2)}{z_j} \geq \frac{f(x_1,x_2)}{z_i}
\right) \, \mbox{d} x_1 \, \mbox{d}  x_2
\Bigg\},
\end{split}
\end{equation*}
where $f(x_1,x_2)$ is the bivariate normal density of $(X_1,X_2)\sim N((0,0)^\top,\bSigma)$, 
and for brevity we set $\bh=(\bt_j - \bt_i)^\top\equiv(t_1,t_2)^\top$. 
Recall from (\ref{Gaussdist}) that
$$
 a(\bh)=(\bh^T\bSigma^{-1} \bh)^{1/2}=\frac{1}{\sqrt{(1 -\rho^2)}}\left(
\frac{t_1^2}{\sigma_1^2}-\frac{2\rho t_1 t_2}{\sigma_1 \sigma_2}
+\frac{t_2^2}{\sigma_2^2}\right)^{1/2}
$$
where $\rho=\sigma_{12}/\sigma_1\sigma_2$.
Consider first the case $(t_1\sigma_2-\rho t_2\sigma_1)>0$. Note that
$f(x_1,x_2)/z_i \geq f(x_1-t_1,x_2-t_2)/z_j$ implies that
\begin{equation*}\label{}
\begin{split}
& \exp\left\{-\frac{1}{2(1-\rho^2)}\left(
\frac{x_1^2}{\sigma_1^2}- \frac{2\rho x_1 x_2}{\sigma_1 \sigma_2} + 
\frac{x_2^2}{\sigma_2^2}\right)\right\}/z_i\\
&\geq
\exp\left\{-\frac{1}{2(1-\rho^2)}\left(
\frac{(x_1-t_1)^2}{\sigma_1^2}-\frac{2\rho (x_1-t_1) (x_2-t_2)}{\sigma_1 \sigma_2} + 
\frac{(x_2-t_2)^2}{\sigma_2^2}\right)\right\}/z_i\\
& \Leftrightarrow x_1 \leq 
\frac{\sigma_1^2 \sigma_2}{2(t_1\sigma_2 -\rho t_2\sigma_1)}\left(
\frac{t_1^2}{\sigma_1^2}-\frac{2\rho t_1 t_2}{\sigma_1 \sigma_2} + 
\frac{t_2^2}{\sigma_2^2}\right)+
\frac{\sigma_1^2 \sigma_2(1-\rho^2)}{t_1\sigma_2 -\rho t_2\sigma_1}
\log\frac{z_j}{z_i}-\frac{x_2(t_2\sigma_1^2 -\rho t_1\sigma_1\sigma_2)}
{t_1\sigma_2 -\rho t_2\sigma_1}\\
&\Leftrightarrow x_1 \leq c.
\end{split}
\end{equation*}
From this, it follows that
\begin{equation*}\label{}
\begin{split}
&\int_{-\infty}^\infty \int_{-\infty}^\infty
\frac{f(x_1,x_2)}{z_i} \mathbb{I}\left(
\frac{f(x_1,x_2)}{z_i} \geq \frac{f(x_1-t_1,x_2-t_2)}{z_j}
\right) \, \mbox{d} x_1 \, \mbox{d} x_2\\
&=\frac{1}{z_i}\int_{-\infty}^\infty \int_{-\infty}^c
\frac{1}{2\pi \sqrt{1-\rho^2}}
\exp\left\{-\frac{1}{2(1-\rho^2)}\left(
\frac{x_1^2}{\sigma_1^2}- \frac{2\rho x_1 x_2}{\sigma_1 \sigma_2} + 
\frac{x_2^2}{\sigma_2^2}\right)\right\}
\, \mbox{d} x_1 \, \mbox{d} x_2\\
&=\frac{1}{z_i}\int_{-\infty}^\infty \varphi(x_2)
\int_{-\infty}^c \varphi\left(\frac{x_1-\rho\sigma_1 x_2/\sigma_2}
{\sigma_1 \sqrt{1-\rho^2}}\right)
\, \mbox{d} x_1 \, \mbox{d} x_2\\
&=\frac{1}{z_i}\int_{-\infty}^\infty \varphi(x_2)
\Phi\left(\frac{c-\rho\sigma_1 x_2/\sigma_2}
{\sigma_1 \sqrt{1-\rho^2}}\right)\mbox{d} x_2\\
&=\frac{1}{z_i}\Phi\left(
\frac{1}{2\sqrt{(1 -\rho^2)}}\left(
\frac{t_1^2}{\sigma_1^2}-\frac{2\rho t_1 t_2}{\sigma_1 \sigma_2}
+\frac{t_2^2}{\sigma_2^2}\right)^{1/2}+
\left(
\frac{t_1^2}{\sigma_1^2}-\frac{2\rho t_1 t_2}{\sigma_1 \sigma_2}
+\frac{t_2^2}{\sigma_2^2}\right)^{-1/2}\frac{\log z_j/z_i}{(1 -\rho^2)^{-1/2}}
\right)\\
&=\frac{1}{z_i}\Phi\left(\frac{a(\bh)}{2}+\frac{\log z_j/z_i}{a(\bh)}\right).
\end{split}
\end{equation*}
Similarly,
$$
\frac{f(x_1-t_1,x_2-t_2)}{z_j} \geq \frac{f(x_1,x_2)}{z_i}
\Leftrightarrow  x_1 \geq c.
$$
It then follows that
\begin{equation*}\label{}
\begin{split}
&\int_{-\infty}^\infty \int_{-\infty}^\infty
\frac{f(x_1-t_1,x_2-t_2)}{z_j} \mathbb{I}\left(
\frac{f(x_1,x_2)}{z_i} \geq \frac{f(x_1-t_1,x_2-t_2)}{z_j}
\right) \, \mbox{d} x_1 \, \mbox{d} x_2\\
&=\frac{1}{z_j}\int_{-\infty}^\infty \int_c^\infty
\frac{1}{2\pi \sqrt{1-\rho^2}}
\exp\left\{-\frac{1}{2(1-\rho^2)}\left(
\frac{(x_1-t_1)^2}{\sigma_1^2}- \frac{2\rho (x_1-t_1) (x_2-t_2)}{\sigma_1 \sigma_2} + 
\frac{(x_2-t_2)^2}{\sigma_2^2}\right)\right\}
\, \mbox{d} x_1 \, \mbox{d} x_2\\
&=\frac{1}{z_j}\int_{-\infty}^\infty \varphi(x_2-t_2)
\int_c^\infty \varphi\left(\frac{(x_1-t_1)-\rho\sigma_1 (x_2-t_2)/\sigma_2}
{\sigma_1 \sqrt{1-\rho^2}}\right)
\, \mbox{d} x_1 \, \mbox{d} x_2\\
&=\frac{1}{z_j}\int_{-\infty}^\infty \varphi(x_2-t_2)
\left\{1-\Phi\left(\frac{c-\rho\sigma_1 (x_2-t_2)/\sigma_2}
{\sigma_1 \sqrt{1-\rho^2}}\right)\right\}\mbox{d} x_2\\
&=\frac{1}{z_j}\Phi\left(
\frac{1}{2\sqrt{(1 -\rho^2)}}\left(
\frac{t_1^2}{\sigma_1^2}-\frac{2\rho t_1 t_2}{\sigma_1 \sigma_2}
+\frac{t_2^2}{\sigma_2^2}\right)^{1/2}+
\left(
\frac{t_1^2}{\sigma_1^2}-\frac{2\rho t_1 t_2}{\sigma_1 \sigma_2}
+\frac{t_2^2}{\sigma_2^2}\right)^{-1/2}\frac{\log z_i/z_j}{(1 -\rho^2)^{-1/2}}
\right)\\
&=\frac{1}{z_j}\Phi\left(\frac{a(\bh)}{2}+\frac{\log z_i/z_j}{a(\bh)}\right),
\end{split}
\end{equation*}
and the form of the distribution (\ref{Gaussdist}) is confirmed.
Observe, that the same result is obtained for the case $(t_1\sigma_2-\rho t_2\sigma_1)<0$. See also \cite{smith90} and \cite{deHaan+p06}.


\subsection*{A.3: Derivation of the bivariate density function}

In order to derive the bivariate density function (\ref{GaussDensity})
we require the second-order derivative of 
$$
F(z_i,z_j)=\exp\left(-\frac{\Phi(w)}{z_i}-
\frac{\Phi(v)}{z_j}\right)
$$
with respect to $z_i$ and $z_j$, where for brevity we
set $a \equiv a(h)$, $w \equiv w(h)$ and 
$v \equiv v(h)$ and write $w=a/2+\log(z_j/z_i)/a$ 
and $v=a-w$. The differentiation gives
\begin{equation*}\label{}
\begin{split}
&f(z_i,z_j)\equiv
\frac{\partial^2}{\partial z_i \partial z_j}F(z_i,z_j)
=\\
&\exp\left(-\frac{\Phi(w)}{z_i}-
\frac{\Phi(v)}{z_j}\right)\Bigg\{
\frac{\partial}{\partial z_i}
\left(-\frac{\Phi(w)}{z_i}-\frac{\Phi(v)}{z_j}\right)
\frac{\partial}{\partial z_j}
\left(-\frac{\Phi(w)}{z_i}-\frac{\Phi(v)}{z_j}\right)
+\frac{\partial^2}{\partial z_i \partial z_j}
\left(-\frac{\Phi(w)}{z_i}-\frac{\Phi(v)}{z_j}\right)
\Bigg\}.
\end{split}
\end{equation*}
First-order differentation gives
$$
\frac{\partial}{\partial z_i}
\left(-\frac{\Phi(w)}{z_i}-\frac{\Phi(v)}{z_j}\right)
=\frac{\Phi(w)}{z_i^2}+\frac{\varphi(w)}{az_i^2}-
\frac{\varphi(v)}{a z_i z_j}, \quad
\frac{\partial}{\partial z_j}
\left(-\frac{\Phi(w)}{z_i}-\frac{\Phi(v)}{z_j}\right)
=\frac{\Phi(v)}{z_j^2}+\frac{\varphi(v)}{az_j^2}-
\frac{\varphi(w)}{a z_i z_j},
$$
using the results
$$
\frac{\partial \Phi(w)}{\partial z_i}=-
\frac{\varphi(w)}{a z_i}, \quad
\frac{\partial \Phi(v)}{\partial z_i}=
\frac{\varphi(v)}{a z_i} \quad \mbox{and}
\quad
\frac{\partial w}{\partial z_i}=-\frac{1}{a z_i},
\quad
\frac{\partial v}{\partial z_i}=\frac{1}{a z_i}. 
$$
%
%
Second-order differentation yields
$$
\frac{\partial^2}{\partial z_i \partial z_j}
\left(-\frac{\Phi(w)}{z_i}-\frac{\Phi(v)}{z_j}\right)
=\frac{v \, \varphi(w)}{a^2 z_i^2 z_j}+\frac{w \, 
\varphi(v)}{a^2 z_i z_j^2},
$$
using
$$
\frac{\partial \varphi(w)}{\partial z_i}=\frac{w \varphi(w)}{a z_i}
\quad \mbox{and} \quad 
\frac{\partial \varphi(v)}{\partial z_i}=-\frac{v \varphi(v)}{a z_i}.
$$
%
%
Substituting, we obtain the probability density function
\begin{equation*}\label{}
\begin{split}
f(z_i,z_j)&=\exp\bigg\{-\frac{\Phi(w)}{z_i}
-\frac{\Phi(v)}{z_j}\bigg\}\Bigg\{
\left(\frac{\Phi(w)}{z_i^2}+\frac{\varphi(w)}{a z_i^2}-
\frac{\varphi(v)}{a z_iz_j}\right)\\
&\bigg(\frac{\Phi(v)}{z_j^2}+\frac{\varphi(v)}{az_j^2}-
\frac{\varphi(w)}{a z_i z_j}\bigg)+
\bigg(\frac{v \, \varphi(w)}{a^2 z_i^2 z_j}+\frac{w \, 
\varphi(v)}{a^2 z_i z_j^2}\bigg)\Bigg\}.
\end{split}
\end{equation*}

\subsection*{A.4: An expression for the squared score statistic}

From Section \ref{sec:MaxStablePL} the term $\text{J}(\bpsi)$ of the
Godambe information matrix can be estimated from
$$
\sum_{i=1}^\mathcal{K} \sum_{j=i+1}^{\mathcal{K}-1} 
\Dlpsi \log f(\by_i,\by_j;\bpsi)^\top \,
\Dlpsi \log f(\by_i,\by_j;\bpsi),
$$
where $\bpsi^\top=(\bsigma,\bbeta_\mu,\bbeta_\lambda,\bbeta_\xi)$, 
 $\bsigma^\top=(\bsigma_{11}^2,\bsigma_{12},\bsigma_{22}^2)$ and where 
each parameter
$\bbeta$ is $p$-dimensional vector of coefficients. 
The bivariate
log-density has the form
$$
\log f(\by_i,\by_j;\bpsi)=\text{A}+ \log(\text{B} \odot \text{C}
+\text{D})+\text{E},
$$
where 
$$
\text{A}=-\frac{\Phi(\bw)}{\bz_i}-\frac{\Phi(\bv)}{\bz_j}, \qquad
\text{B}=\frac{\Phi(\bw)}{\bz_i^2}+
\frac{\varphi(\bw)}{\ba \odot \bz_i^2}-
\frac{\varphi(\bv)}{\ba \odot \bz_i \odot \bz_j}, \, 
$$
$$
\text{C}=\frac{\Phi(\bv)}{\bz_j^2}+
\frac{\varphi(\bv)}{\ba \odot \bz_j^2}-
\frac{\varphi(\bw)}{\ba \odot \bz_i \odot \bz_j}, \qquad
\text{D}=\frac{\bv \odot \varphi(\bw)}{\ba^2 \odot \bz_i^2 \odot \bz_j}+
\frac{\bw \odot \varphi(\bv)}{\ba^2 \odot \bz_i \odot \bz_j^2}\,
$$
$$
\mbox{and} \quad  \text{E}=\log\left\{ \frac{1}{\lambda_i 
\lambda_j}\left(\bone + \xi_i
\frac{\by_i-\mu_i \bone}{\lambda_i}\right)_+^{\frac{1}{\xi_i}-1}
\left(\bone + \xi_j
\frac{\by_j-\mu_j \bone }{\lambda_j}\right)_+^{\frac{1}{\xi_j}-1}
\right\}
$$
and where $\bone^\top=(1, \ldots, 1)$, $\mu_i=(\bX_{\sbbeta_\mu}\bbeta_\mu)_i$, 
$\xi_i=(\bX_{\sbbeta_\xi}\bbeta_\xi)_i$ and 
$\log(\psi_i)=(\bX_{\sbbeta_\psi}\bbeta_\psi)_i$. 
The GEV parameters are related to the 
predictors by the form (\ref{eq:Predict}).
We assume identity link functions 
for the location and shape parameters and exponential 
for the scale. The term $\text{E}$ corresponds to the log of the
 determinat of the Jacobian matrix associated with the transformation 
(\ref{eq:GEVTransf}), see Section \ref{sec:MaxStablePL}.  

The first-order derivative term of the square score statistic
is defined by
$$\Dif_{\sbpsi}\log f(\by_i,\by_j;\bpsi)=(\Dif_{\bsigma} 
\log f(\bpsi),\Dif_{\sbbeta_\mu} \log f(\bpsi), 
\Dif_{\sbbeta_\lambda} \log f(\bpsi), \Dif_{\sbbeta_\xi} \log f(\bpsi))$$ 
where for brevity we write $\log f(\bpsi)\equiv\log f(\by_i,\by_j;\bpsi)$. 
Vector differential calculus (e.g. \cite{wand02}) leads to 
\begin{equation*}\label{}
\begin{split}
\Dif_{\sigma}\log f(\bpsi)&\equiv 
\Bigg[-\frac{\bv \odot \varphi(\bw) }{\ba \odot \bz_i}-
\frac{\bw \odot \varphi(\bv) }{\ba \odot \bz_j}
+\bigg\{\text{C} \odot \bigg(\frac{(\bw^2-\bone) \odot \varphi(\bv)}
{\ba^2 \odot \bz_j^2}+\frac{(\bone+\bw \odot \bv) \odot \varphi(\bw)}
{\ba^2 \odot \bz_i \odot \bz_j}\bigg)\\
&+
\text{B} \odot \bigg(\frac{(\bv^2-\bone) \odot \varphi(\bw)}
{\ba^2 \odot \bz_i^2}
+\frac{(\bone+\bw \odot \bv) \odot \varphi(\bv)}
{\ba^2 \odot \bz_i \odot \bz_j}\bigg)+
\frac{(\bw-2\bv-\bw \odot \bv^2) \odot 
\varphi(\bw)}{\ba^3 \odot \bz_i^2 \odot \bz_j}\\
&+
\frac{(\bv-2\bw-\bv \odot \bw^2) \odot \varphi(\bv)}
{\ba^3 \odot \bz_j^2 \odot \bz_i}\bigg\}
/(\text{B} \odot \text{C}+\text{D})\Bigg]\bs^\top,
\end{split}
\end{equation*}
where $\bs^\top\equiv (t_1^2, \, 2 \,t_1 \, t_2, \,t_2^2)$, using the results
%
$$
\Dif_{\bsigma} \bw = \frac{\bv}{\ba} \, \bs^\top, \qquad
\Dif_{\bsigma} \Phi(\bw) = \frac{\bv \odot \varphi(\bw)}{\ba} \, \bs^\top, \qquad
\Dif_{\bsigma} \varphi(\bw) = - \frac{\bw \odot \bv\varphi(\bw)}{\ba} \, \bs^\top
$$
$$
\mbox{and} \quad \Dif_{\bsigma} \bv \odot \varphi(\bw) = 
\frac{\bw \odot (\bone-\bv^2)\varphi(\bw)}{\ba} \, \bs^\top. 
$$
The first-order derivativies of $\bv$, 
$\Phi(\bv)$, $\varphi(\bv)$ and $\bw \odot \varphi(\bv)$ are 
the same as the above, substituting $\bv$ for $\bw$.
Similarly for the second term we have
\begin{equation*}\label{}
\begin{split}
\Dif_{\sbbeta_\mu} \log f(\bpsi) &\equiv
\Bigg\{\bigg(\frac{\varphi(\bw)+\ba \odot 
\Phi(\bw)}{\ba \odot \bz_i^2}
-\frac{\varphi(\bv)}{\ba \odot \bz_i \odot \bz_j}\bigg) \odot 
\frac{\bz_i^{1-\xi_i}}{\lambda_i} 
\Bigg\}(\bX_{\sbbeta_\mu})_i\\
&+
\Bigg\{\bigg(\frac{\varphi(\bv)+\ba \odot \Phi(\bv)}{\ba \odot \bz_j^2}
-\frac{\varphi(\bw)}{\ba \odot \bz_i \odot \bz_j}\bigg) \odot  
\frac{\bz_j^{1-\xi_j}}{\lambda_j}\Bigg\}
(\bX_{\sbbeta_\mu})_j\\
&+
\left[ \frac{\bigg\{ \text{C} \odot \bigg(\frac{\bw \odot 
\Phi(\bv)}{\ba^2 \odot \bz_i \odot \bz_j^2}+
\frac{\bv \odot \Phi(\bw)}{\ba^2 \odot \bz_j \odot \bz_i^2}\bigg) 
\odot \frac{\bz_i^{1-\xi_i}}{\lambda_i}
\bigg\}}{(\text{B} \odot \text{C}+\text{D})}
\right](\bX_{\sbbeta_\mu})_i\\
&+\left[ \frac{\bigg\{ \text{C}\odot \bigg(\frac{(\ba+\bw) \odot 
\varphi(\bw)}{\ba^2 \odot \bz_i \odot \bz_j^2}-
\frac{(2\ba+\bw) \odot \varphi(\bv)}{\ba^2 \odot \bz_j^3}-
\frac{2\Phi(\bv)}{\bz_i^3}\bigg) \odot 
\frac{\bz_j^{1-\xi_j}}{\lambda_j}\bigg\}}{(\text{B} \odot 
\text{C}+\text{D})} \right]  (\bX_{\sbbeta_\mu})_j \\
&+
\left[ \frac{\bigg\{ \text{B} \odot \bigg(\frac{(\ba+\bv) \odot 
\varphi(\bv)}{\ba^2 \odot \bz_j \odot \bz_i^2}-
\frac{(2\ba+\bv) \odot \varphi(\bw)}{\ba^2 \odot \bz_i^3}-
\frac{2\Phi(\bw)}{\bz_j^3}\bigg) 
\odot \frac{\bz_i^{1-\xi_i}}{\lambda_i}\bigg\}}{
(\text{B} \odot \text{C}+\text{D})} 
\right] (\bX_{\sbbeta_\mu})_i\\
&+\left[\frac{\bigg\{\text{B} \odot \bigg(\frac{\bv \odot \Phi(\bw)}
{\ba^2 \odot \bz_j \odot \bz_i^2}+\frac{\bw \odot \Phi(\bv)}
{\ba^2 \odot \bz_i \odot \bz_j^2}\bigg) \odot \frac{\bz_j^{1-\xi_j}}{\lambda_j}
\bigg\}}{(\text{B} \odot \text{C}+\text{D})} \right](\bX_{\sbbeta_\mu})_j\\
&+\left[ \frac{\bigg\{ \bigg(\frac{(\bone-\ba \odot \bv-\bv^2) \odot 
\varphi(\bw)}{\ba^2 \odot \bz_j \odot \bz_i^3}-
\frac{(\bone-\ba \odot \bw-\bw^2) \odot \varphi(\bv)}
{\ba^2 \odot \bz_i \odot \bz_j^3}\bigg) \odot 
\frac{\bz_i^{1-\xi_i}}{\lambda_i}\bigg\}}{(\text{B} \odot \text{C}+\text{D})}
\right](\bX_{\sbbeta_\mu})_i\\
&+\left[ \frac{\bigg\{\bigg(\frac{(\bone-\ba \odot \bw-\bw^2) \odot \varphi(\bv)}
{\ba^2 \odot \bz_i \odot \bz_j^3}-
\frac{(\bone-\ba \odot \bv-\bv^2) \odot \varphi(\bw)}
{\ba^2 \odot \bz_j \odot \bz_i^3}\bigg)\odot 
\frac{\bz_j^{1-\xi_j}}{\lambda_j}\bigg\}}{(\text{B} \odot \text{C}+\text{D})}
\right] (\bX_{\sbbeta_\mu})_j\\
&+\frac{(\xi_i-1)}{\lambda_i \bz^{\xi_i}} (\bX_{\sbbeta_\mu})_i
+\frac{(\xi_j-1)}{\lambda_j \bz^{\xi_j}} (\bX_{\sbbeta_\mu})_j.
\end{split}
\end{equation*}
Observe that the above expression is obtained by 
deriving in order the components: 
$\Dif_{\sbbeta_\mu} \text{A}$,
$\Dif_{\sbbeta_\mu} \log(\text{B} \odot \text{C}+\text{D})$ and 
$\Dif_{\sbbeta_\mu} \text{E}$. These three 
components have the form $\Dif_{\sbbeta_\mu} f(\bx)=
\Dif_{\bz_i} f(\bx) \Dif_{\sbbeta_\mu} \bz_i+
\Dif_{\bz_j} f(\bx) \Dif_{\sbbeta_\mu} \bz_j$, where 
$\Dif_{\sbbeta_\mu} \bz_i=\bz_i^{1-\xi_i}/\lambda_i \bX_{\sbbeta_\mu}$. 
For this reason the derived expressions of $\Dif_{\sbbeta_\lambda} \log f(\bpsi)$
and $\Dif_{\sbbeta_\xi} \log f(\bpsi)$ are essentially the same but 
substituting $\Dif_{\sbbeta_\mu} \bz_i$ with $\Dif_{\sbbeta_\lambda} \bz_i$
and $\Dif_{\sbbeta_\xi} \bz_i$, and $\Dif_{\sbbeta_\mu} \text{E}$ with
$\Dif_{\sbbeta_\lambda} \text{E}$ and $\Dif_{\sbbeta_\xi} \text{E}$.
We have
$$
\Dif_{\sbbeta_\lambda} \bz_i \equiv - \frac{\bz_i (\by_i-\mu_i \bone)}{\lambda_i}
(\bX_{\sbbeta_\lambda})_i \quad \mbox{and} \quad
\Dif_{\sbbeta_\xi} \bz_i \equiv \left\{\frac{1}{\xi_i}
\left(\frac{\bz_i^{1-\xi_i} (\by_i-\mu_i \bone)}
{\lambda_i}-\bz_i \log(\bz_i)\right)\right\}
(\bX_{\sbbeta_\xi})_i.
$$
%
Finally,
$$
\Dif_{\sbbeta_\lambda} \text{E} \equiv \left( \frac{(\xi_i-1)
(\by_i - \mu_i \bone)}{\lambda_i \bz_i} - 
\bone \right)(\bX_{\sbbeta_\lambda})_i+
\left( \frac{(\xi_j-1)
(\by_j - \mu_j \bone)}{\lambda_j \bz_j} - 
\bone \right)(\bX_{\sbbeta_\lambda})_j 
\quad \mbox{and}
$$
\begin{equation*}\label{}
\begin{split}
\Dif_{\sbbeta_\xi} \text{E} 
&\equiv \left[ \frac{1}{\xi_i}
\left\{ \frac{(1-\xi_i)(\by_i-\mu_i \bone)}{\bz_i^{\xi_i}\lambda_i}
-\log(\bz_i) \right\}\right](\bX_{\sbbeta_\xi})_i\\
&+ \left[ \frac{1}{\xi_j} 
\left\{ \frac{(1-\xi_j)(\by_j-\mu_j \bone)}{\bz_j^{\xi_j}\lambda_j}
-\log(\bz_j) \right\}\right](\bX_{\sbbeta_\xi})_j.
\end{split}
\end{equation*}
In combination we obtain an expression for
$\Dif_{\sbpsi}\log f(\by_i,\by_j;\bpsi)$, and from this the 
squared score statistic.

\end{document}